# FREQUENCY STABILITY OF ATOMIC FREQUENCY STANDARDS BEYOND QUANTUM PROJECTION NOISE

G.M.SAXENA*

Time and Frequency Section
National Physical Laboratory
Dr K.S.Krishnan Road, New Delhi-110012 India

*E-mail gmsaxena@mail.nplindia.ernet.in
Ex-Sr Scientist

**Abstract:** In this paper we describe that the optically pumped frequency standards can have frequency stability beyond the quantum noise limit by detecting the Ramsey resonance through the squeezed light. In this paper we report that instead of considering the interaction of entangled atoms in the microwave region, it will be more practical to create the entanglement of the atoms in the detection region using the squeezed light, which is also used for the detection of the Ramsey resonance. The advantage of squeezing can be derived when the technical noises have been removed.

**Introduction:** The interaction of the squeezed states of light with the two level atoms has been extensively studied in the last two decades. The attention has been focussed in the recent years on the applications of the squeezed states in practical devices. The atomic Cs beam and the fountain frequency standards provide the scope for the application of the squeezed states in suppressing the atomic shot noise. Recently, it has been reported that the fountain frequency stability reached the limit of the atomic shot or projection noise[1]. Several studies have been made on the suppression of the projection noise. The possible solution lies in creation of the entangled states of the atoms and then using these atoms for the detection of the Ramsey correction signal with improved S/N. However, the creation of the atomic entangled states is a formidable task. Some of the researchers have suggested the realization of the entangled atomic states through the squeezed light. As it is considered that low noise characteristics, in one of the quadratures of the squeezed light, may be transferred to the atomic assembly under certain conditions. And we may obtain higher S/N. In the atomic frequency standards, it is expected that the entangled states of atoms may improve the S/N. In this paper we describe the application of the squeezed states for the entanglement of the atoms in the detection region in the Cs Time and Frequency standard. This idea is practical in the sense that in some of the optically pumped atomic clocks, the correction signal is obtained through the resonance fluorescence signal. The resonance fluorescence signal, from the two level atoms interacting with the squeezed vacuum coloured by coherent states, shows reduced fluctuations in atomic spin in one of the quadratures. The four-pi problem is to be taken care of in such a detection scheme. It is observed that detecting the resonance fluorescence with a special type of the homodyne detector[2] can surmount this problem. It is assumed that there is negligible absorption of light by the atoms.

It has also been shown theoretically that the decay of the atoms to the pure state may eliminate the projection noise. In the hyperfine two-level ground state of Cs atoms, generation of the pure state can be realized through the squeezed states[3]. The state is related to the correlated atomic states. Putting the atoms in the pure state will result in resonance signal with higher S/N.

## Cesium Beam and Fountain Frequency Standards:

Let us consider the case in which almost all of the atoms have been put in the ground state (g.s) hyperfine level F=4, $m_F$ =0 after the interaction of the Cs atoms with the microwave field in the Ramsey cavity configuration either in Cs beam[4] or the fountain clock[5]. The correction signal is obtained by detecting the resonance fluorescence signal for the transitions between F=4 ,$m_F$ =0 – F'=5, $m_{F'}$ =0, ±1. The fluorescent signal is the measure of the atoms undergoing the clock transitions. We consider irradiation of the Cs atoms with the resonant squeezed vacuum and the coherent radiation for obtaining the resonance fluorescence from F=4, $m_F$ =0 - F'=5, $m_{F'}$ =0,±1, transitions. With proper phase matching conditions we can detect the resonance fluorescence in the quadrature having reduced quantum noise. We shall discuss the situation in which atomic projection noise is below the zero point fluctuations in the detected resonance fluorescence. It has already been established that squeezed vacuum coloured by the coherent resonant light reduces the atomic spin noise. However, much depends on the phase matching conditions. We consider the case in which the time of spontaneous decay in the reduced quantum noise quadrature is larger than the interaction time between the atoms and the squeezed light in the detection region. The contribution of the projection noise to the noise spectra in the detection scheme of the ref. 2 is proportional to $N_{AT}(\cos^2 \vartheta -1)\cos 2\varphi$, here $N_{AT}$ is the atomic flux, $\vartheta$ is the Bloch angle rotation caused by the driving field and $\varphi$ is the relative phase between the local oscillator and driving field. On subjecting the atoms in the g.s F=3, $m_F$=0 to the r-f radiation in the Ramsey separated cavity configuration, the atoms are pumped to F=4, $m_F$=0 after the rf transitions. The probability of finding the atoms in the F=4, $m_F$=0 state is nearly one after the microwave interaction. At this point, for obtaining the correction signal we shine on the atoms the coherent squeezed light which is resonant with the F=4,$m_F$=0 –F'=5,$m_{F'}$=0, ±1 transitions.

As shown in the Figs.1a and 1b the electric field of the applied detection light is along the x-axis. The atomic magnetic dipole is also rendered parallel to x-axis by applying weak magnetic field along the x-axis. This also serves the purpose of the magnetic field required for the removal of the degeneracy in the h.f levels. As F=4 – F' =5 is the cyclic transition, the resonance fluorescence will be due to the atoms which occupy g.s F=4, $m_F$=0 level. It is also because the decay into one of the quadratures is made slower by suitably choosing the N and M parameters which describe the number of photons and their correlation respectively in the squeezed light. These parameters basically define the squeezing characteristics of the light. The resonance fluorescence signal detected at photo- detector 1 (Fig.-1a and Fig.-1b) has narrowed spectrum due to the use of the squeezed light. The direction of the electric vector of the field may be fixed by the polarizer (P1). The relative phase between the driving field and the local oscillator is adjusted by the Babinet compensator (SB). With the suitable choice of the phase, using the Babinet compensator (SB), the desired resonance signal may be detected in the quadrature having reduced quantum noise. We shall now discuss the cases with coherent and squeezed light as detection light respectively. The S/N with coherent light is expressed as

$$S/N_{VAC} = N\Delta^2/2, \qquad (1)$$

here $N = \alpha T(\beta|A|)^2/B$ and $\Delta$ is the atomic response for the fields[6]. When the coherent light is replaced by squeezed light, the signal to noise ratio is given by the expression,

$$S/N_{SQ} = N\Delta^2_{SQ}/[2\{1 +\xi\, S(\Omega_0, \varphi_-)\}]. \qquad (2)$$

$S(\Omega_0, \varphi_-)$ is the spectrum of the squeezing. It includes the contribution of the atomic projection noise, which is detected in the quadrature having reduced quantum noise. As the atomic projection noise happens to dominate we ignore the terms arising from the fluctuations in the mean dipole moment and spontaneous emission noise. $\Delta_{SQ}$ is the atom's response for the squeezed field. It is independent of the phase. The noise characteristics of the interacting field are transferred to the atoms through $\Delta_{SQ}$. On careful examination of the above two equations we find that the squeezed light introduces phase dependence in the quantum noise distribution. The spectrum $S(\Omega_0, \varphi_-)$ for the squeezed states may be written as[2]

$$S(\Omega_0, \varphi_-) = C(\eta_S, \eta_0, \acute{N}_{AT}, P_{LO}, P_x) F(\Omega_0) \cos 2\varphi_-, \qquad (3)$$

in the above expression C is a function of $P_x$ the power of squeezed driving field, $P_{LO}$ the power of the local oscillator and $\eta_S$ & $\vartheta_M$ are the efficiencies ref.2. $F(\Omega_0)$ is the term which defines the impact of the noise characteristics of the field on the atomic system. Its value depends on the experimental conditions. $\acute{N}_{AT}$ is the rate at which the Cs atoms cross the interaction region or the atomic flux. The term $F(\Omega_0)\cos 2\varphi_-$ is the phase dependent part of the resonance fluorescence. By suitably choosing the squeezed light parameters and phase between the squeezed light and driving field, we can obtain the squeezed noise spectrum having reduced quantum fluctuations. For the atoms interacting with the squeezed coherent light, the resonance fluorescence spectrum will show the narrowing of the resonance fluorescence spectrum and tends to negative values thus resulting in higher S/N[7]. From the eqs.2 and 3 we can obtain the S/N of the signal for the squeezed driving field. Thus we can obtain the better stability with the squeezed light. The frequency stability of an optically pumped frequency standard is given by[8]

$$\sigma(\tau) \approx \frac{\delta\upsilon}{\upsilon (S/N)} \tau^{-1/2}, \qquad (4)$$

here $\upsilon$ is the microwave clock transition frequency (9.192 G$_Z$H), $\delta\upsilon$ is its linewidth, $\tau$ is the sampling time and S/N is the clock signal to noise ratio. The clock S/N represents the effect of all noise processes including the atomic shot noise, laser frequency fluctuations, photon shot noise and the noise of the detection system. As we have mentioned above the atomic projection noise dominates over the other noises. For the squeezed light the signal to noise ratio is given by eqn.2.

**Conclusion :** In this brief paper we have discussed the phase dependent resonance fluorescence spectrum obtained from the Cs atom- squeezed light interaction. The resonance fluorescence signal is used for obtaining the clock correction signal. We have discussed the application of the squeezed light, in the detection region, for reducing the atomic projection noise in the detection quadrature. The degree of suppression of the noise for the practical device like Cs beam or fountain clock can be worked out by incorporating the actual experimental parameters in the eqs.2-3. The advantage of using squeezed light lies in making the results independent of the velocity of the atoms to some extent. It is because the decay time in the less noisy quadrature could be made larger than the natural or spontaneous emission time. The S/N of the clock signal can be increased with suppressed quantum noise and reduced atomic projection noise as the squeezing properties of the detection light are transferred to the atoms. The experimental realization of squeezed spin states of laser cooled atoms through squeezed light has also been reported[9]. It is expected that with technical noise eliminated, atomic frequency standards in the realm of reduced quantum noise will be realizable in near future. In a recent paper[10] we have shown that the squeezed states may also enhance the laser cooling force in fountain atomic clocks. Thus the squeezed states may be very useful in several ways in Time and frequency metrology.

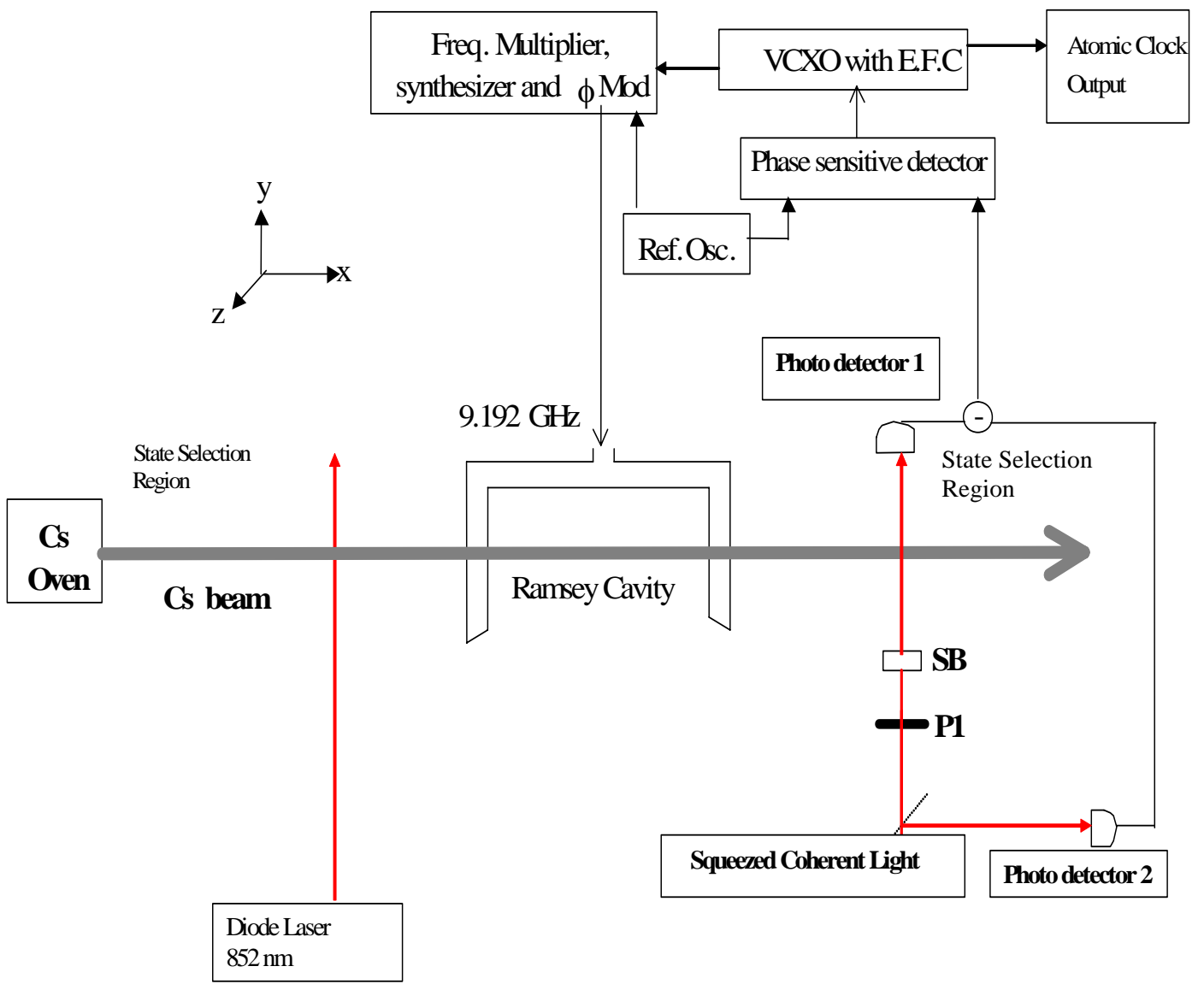

**Fig-1a Schematic of Physics Package of squeezed Cs beam Frequency Standard**

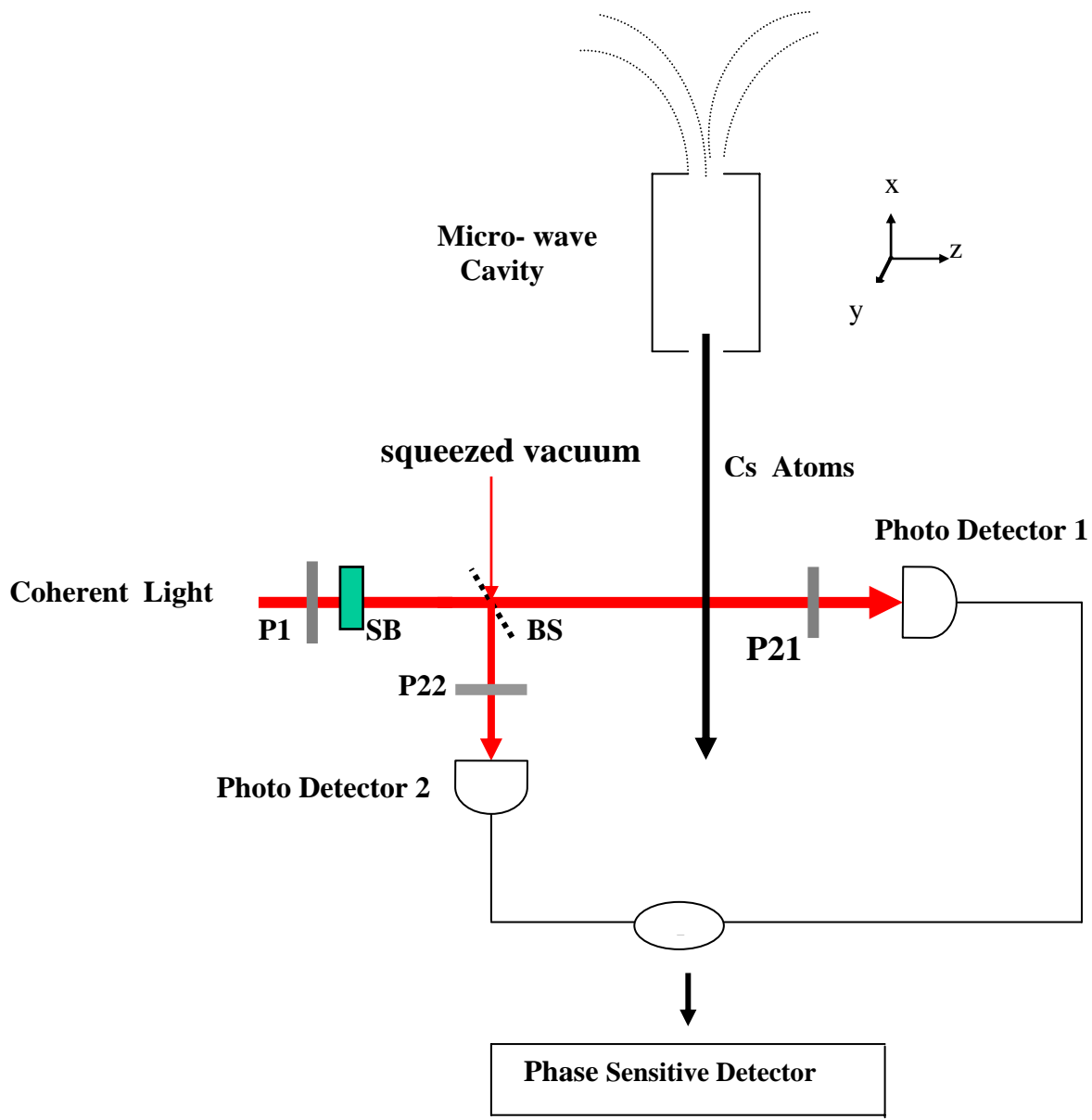

**Fig -1b Schematic of Physics Package of Squeezed Fountain Frequency Standard**